\documentclass[12pt,preprint]{aastex}

\newcommand{\rsun}{\ensuremath{R_\sun}}
\newcommand{\msun}{\ensuremath{M_\sun}}

\newcommand{\ms}{m s$^{-1}$}

\newcommand{\vcb}{$V_{\rm CB}$}
\newcommand{\vcbsun}{$V_{\rm CB, \sun}$}

\newcommand{\figr}[1]{Fig.~\ref{fig:#1}}
\newcommand{\secr}[1]{\mbox{Section \ref{sec:#1}}}
\newcommand{\eqr}[1]{Eq.~\ref{eq:#1}}

\newcommand{\tabr}[1]{\mbox{Table~\ref{tab:#1}}}

\slugcomment{Submitted to ApJ: Feb 4, 2011. Accepted: Mar 15, 2011}

\shorttitle{The impact of the CB effect on spectroscopic planetary transits}

\shortauthors{Shporer \& Brown}

\begin{document}

\title{THE IMPACT OF THE CONVECTIVE BLUESHIFT EFFECT ON SPECTROSCOPIC PLANETARY TRANSITS}

\author{Avi Shporer\altaffilmark{1,2}, 
Tim Brown\altaffilmark{1,2}}

\altaffiltext{1}{Las Cumbres Observatory Global Telescope Network, 6740 Cortona Drive, Suite 102, Santa Barbara, CA 93117, USA; ashporer@lcogt.net}
\altaffiltext{2}{Department of Physics, Broida Hall, University of California, Santa Barbara, CA 93106, USA}

\begin{abstract}

We present here a small anomalous radial velocity (RV) signal expected to be present in RV curves measured during planetary transits. This signal is induced by the convective blueshift (CB) effect --- a net blueshift emanating from the stellar surface, resulting from a larger contribution of rising hot and bright gas relative to the colder and darker sinking gas. Since the CB radial component varies across the stellar surface, the light blocked by the planet during a transit will have a varying RV component, resulting in a small shift of the measured RVs. The CB-induced anomalous RV curve is different than, and independent of, the well known Rossiter-McLaughlin (RM) effect, where the latter is used for determining the sky-projected angle between the host star rotation axis and the planet's orbital angular momentum axis. The observed RV curve is the sum of the CB and RM signals, and they are both superposed on the orbital Keplerian curve. If not accounted for, the presence of the CB RV signal in the spectroscopic transit RV curve may bias the estimate of the spin-orbit angle. In addition, future very high precision RVs will allow the use of transiting planets to study the CB of their host stars.

\end{abstract}


\section{Introduction}
\label{sec:intro}

Radial velocity (RV) measurements of Sun-like stars are now reaching the 1 \ms\ accuracy milestone and are expected to become more accurate in the future \citep[e.g.,][]{li08, dumusque10, howard10, locurto10}. Naturally, the increased sensitivity allows to identify more and more low amplitude effects that need to be accounted for when modeling the measured RV curves. We present here such an effect, induced by what is known as stellar convective blueshift (hereafter CB; explained in more detail below), which manifests itself as a small anomalous RV signal during a planetary transit. 

This small RV signal is superposed on, and independent of, the already well known Rossiter-McLaughlin (RM) effect (\citealt{rossiter24}; \citealt{mclaughlin24}; see also \citealt{holt1893}), seen during a planetary transit \citep[e.g.,][]{queloz00, winn05}. The RM effect originates from the host star rotation, and allows the measurement of $\lambda$, the sky-projected angle between the host star's rotation axis and the planet's orbital angular momentum\footnote{Some authors use $\beta = -\lambda$ to mark this angle.}, which is an important piece of information in the study of planetary formation and orbital evolution 
\citep[e.g.,][]{fabrycky07, triaud10, winn10}. During transit the planet blocks the light coming from different regions of the host star surface, with a different RV component due to the star's rotation. The observed line profile is distorted, it becomes asymmetric, and its center is shifted, resulting in the RM RV signal. The RM RV curve shape depends primarily on the spin-orbit angle $\lambda$, the sky-projected rotation velocity $V_{rot}$, and the planet to star radii ratio $r \equiv R_p/R_s$ \citep{gaudi07}.

The new effect we present here is expected to be at the 1 \ms\ level and to occur during transit, thus affecting the interpretation of the RV curve. We describe the CB effect in \secr{cb}, and in \secr{model} we use a simple numerical model to calculate the impact of the CB effect on RVs measured during transit. We give a discussion in \secr{dis} and a summary in \secr{sum}.

\section{The Convective Blueshift (CB) Effect}
\label{sec:cb}

The history of the convective blueshift effect dates back to when astronomers started investigating in detail the spectrum of the Sun, and identified disagreements between the measured line positions and their expected positions based on lab experiments (\citealt{jewell1896}; for a review see \citealt{dravins82}). We know today that several physical processes contribute to these disagreements, one of them being the relativistic gravitational redshift \citep[e.g.,][]{earman80}. Still, even after the gravitational redshift and the Doppler shift induced by the rotation of the Sun and the Earth were accounted for the above discrepancy was not completely resolved. The corrected spectra showed that spectra of the Sun observed close to the Sun's limb are redshifted relative to spectra from the center of the Sun's disk, and this redshift scales with $\mu\equiv\cos\theta$, where $\theta$ is the angle between the normal to the Sun's surface and the line of sight. This unexpected trend was studied by many \citep[e.g.,][]{evershed36, forbes62, adam76} and its origin was a puzzle for the better part of the 20th century \citep[for an early historic review see][]{forbes61}. Before it was understood it was named the solar redshift problem, or the solar limb effect.

The solution to the problem came from the realization that the surface of a Sun-like star is composed of convective cells, or granules, where hot gas is rising upward, and cooler gas is sinking downward in the intragranular lanes. Therefore, the stellar surface is composed of local blueshifts and redshifts. Since the rising stream of hot gas is brighter than the cooler sinking gas, a net blueshift is observed \citep[e.g.,][]{beckers78, dravins82}. Following the understanding of the physical process responsible for the effect it was referred to as convective blueshift. We denote the ``local" blueshift velocity as \vcb, which is the blueshift observed at the center of the star's disk, when averaging across a surface area larger than the size of a granulation cell. We note that the \vcb\ velocity is negative. Depending on the value of this velocity, the spectral lines are broadened in an asymmetric way and their center is shifted. For the Sun \vcbsun\ $\approx -300$ \ms\ \citep[e.g.,][]{dravins87}, and it ranges from about $-200$ \ms\ for K-type stars to $-1000$ \ms\ for F-type stars (\citealt{dravins90a}; \citealt{dravins90b}; \citealt{dravins99}). For comparison the Solar gravitational redshift velocity at infinity is $+636$ \ms\ \citep{lindegren03}. Therefore, the CB effect should be accounted for when determining a star's absolute RV, relative to its center of mass velocity \citep{allende02, lindegren03, allende09, delacruz11}, and it should be taken into account when measuring the relative RV between the two components of a binary star, when the required accuracy is better than 1000 \ms\ \citep{pourbaix02}.

Our goal here is to quantify the impact of the CB effect on Doppler shift measurements during a planetary transit. This effect was already briefly mentioned by \cite{winn05}, but its influence on the measured RVs was not studied.

\section{The Model}
\label{sec:model}

We adopt a model where the planet is a spherical dark body, and the host star is a uniformly rotating spherical body with a limb darkened surface. We further assume that the host star has a uniform CB effect, where the radiation emanating from each position on its surface is blueshifted by \vcb\  directed perpendicular to the surface. A hidden assumption here is that the transiting planet size (or the stellar disk area the planet transverses during a single exposure) is much larger than the typical size of a granular cell, where for the Sun the latter has a diameter of typically up to about 1,000 km \citep[e.g.,][]{roudier87}.

We note that the adopted model is somewhat simplified as it is meant to present a first step at understanding the impact of the CB effect on spectroscopic transit observations. The model ignores possible more subtle effects \citep[e.g.,][]{lindegren03, delacruz11}, such as meridional flows on the stellar surface \citep[e.g.,][]{dravins99}, a possible different \vcb\ for different spectral lines \cite[e.g.,][]{dravins82}, or at different stellar latitudes \citep{beckers80}, or a dependence of the local observed blueshift on $\mu$ \citep[e.g.,][see also \secr{dis}]{howard80}. 

In our model the blueshift radial component of each surface element is $\mu$\vcb, so the absolute RV {\it decrease} induced by the CB effect is:
\begin{equation}
\label{eq:rvzp}
V_{0,\rm CB} = \frac{\int_0^1 V_{\rm CB} I(\mu) \mu^2 d\mu}{\int_0^1\! I(\mu) \mu d\mu} = fV_{\rm CB}\ ,
\end{equation}
where the integration is across the stellar hemisphere seen by the observer, $I(\mu)$ is the limb darkening law, and $f$ is an order of unity coefficient. Since in our model \vcb\ is a constant we get:
 \begin{equation}
\label{eq:f}
f = \frac{\int_0^1\! I(\mu) \mu^2 d\mu}{\int_0^1\! I(\mu) \mu d\mu} \ ,
\end{equation}
 which equals 2/3 when there is no limb darkening. As the granulation pattern constantly changes, the value of $f$ will vary with a time scale of a granulation cell lifetime, which for the Sun is about 10 min \citep{asplund00}. However, even for short exposure times this variation is small if not negligible since it is averaged out by the integration across the stellar surface, containing many cells. Although granulation cell characteristics may change significantly for giant stars those are not targeted by high precision RV measurements. We note that the corresponding integral for the RM effect is zero when the planet is out-of-transit, meaning the origin of the RM effect --- the host star rotation --- does not affect the absolute RV, as expected. 

As an example, we have adopted an imaginary star-planet system whose parameters are typical for a transiting system with a close-in giant planet, and are listed in \tabr{params}. We used a planetary radius and mass of $R_p = 1.2\ R_J$ and $M_p = 0.8\ M_J$, and assumed a 3.5 day orbit which is completely edge-on ($i$=90 deg), circular ($e$=0) and aligned ($\lambda$=0 deg). For the host star we assumed it is a G0 type star with radius and mass of $R_s = 1.10\ \rsun$ and $M_s = 1.05\ \msun$ \citep{cox00}, rotational velocity of $V_{rot}$ = 2,000 \ms\ and CB velocity of \vcb\ = $-500$ \ms. We calculated the Keplerian orbit of such a system, and at each phase during the transit we numerically integrated over the visible and eclipsed stellar surfaces in order to calculate both the RM and CB RV shifts. We assumed a quadratic limb darkening law and determined the coefficients using the grids of \cite{claret04} for the SDSS-g' band, where the host star emits most of its radiation. In this model the coefficient $f$ in \eqr{rvzp} is 0.72 and $V_{0,\rm CB}$ = $-360$ \ms.

The resulting RV curves are plotted in \figr{cb}. The top panel shows the RV curve during transit induced by the CB effect (blue), while the RM and Keplerian signals are ignored. The middle panel shows the CB RV curve (blue) and the RM RV curve (green) and their sum (black), while ignoring the orbital RV signal. For completeness we show in the bottom panel the CB and RM RV curves including the Keplerian orbit, with the same color code as the middle panel.

\section{Discussion}
\label{sec:dis}

During transit the planet blocks light coming from different regions of the host star's surface. This changes the integrated surface area in \eqr{rvzp}, and since surface elements with different sky-projected distance from the star center have a different RV component (following a different value for $\mu$), the result of the integration varies, giving an anomalous RV signal during transit. This signal is superposed on the Keplerian and RM signals. We note that similarly to the RM effect, the RV variation originates from a variation in the spectral line centers following a variation in their shape, and not a shift of the lines as a whole, which is the case in orbital motion. Since the CB RV shift depends only on $\mu$ (and \vcb, which is constant) the CB RV curve is a {\it symmetric} function about the mid transit time (for circular orbits).

\begin{figure}
\begin{center}
\includegraphics[scale=0.70]{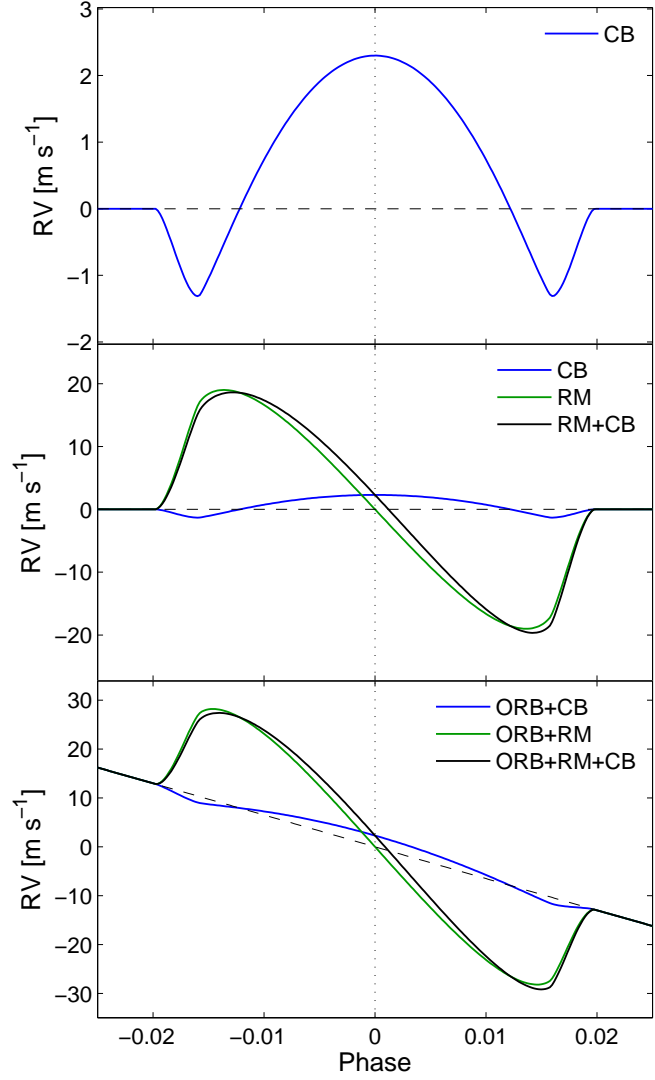}
\caption{\label{fig:cb} 
{\it Top:} The CB-induced RV curve during transit (blue), while the RM and Keplerian signals are ignored. {\it Middle:} The CB RV curve (blue) and RM RV curve (green) and their sum (black), while ignoring the Keplerian signal. {\it Bottom:} Similar to the middle panel, with the Keplerian orbit RV added to all three curves.
All panels show RV in \ms\ versus orbital phase from mid transit time.
}
\end{center}
\end{figure}

As shown by \figr{cb} top panel, the CB effect during transit induces a small anomalous RV signal, at the 1 \ms\ level, although it could be larger for other stars with larger \vcb. The signal reaches maximum at mid transit time since in our model the observed blueshift velocity is largest close to the center of the stellar disk, which is also where most of the light coming from the star is emitted due to limb darkening.  

The reason for the sign change in the CB RV curve can be understood as follows. During the transit start and end the planet blocks light coming from the stellar limb, where the radial blueshift velocity component is smaller than the average across the entire stellar disk. This in turn {\it increases} the overall blueshift, causing a {\it decrease} in the measured velocity. When the planet approaches the stellar disk center, the blueshift RV component of the eclipsed stellar surface {\it increases}, thereby {\it decreasing} the average blueshift across the rest of the disk, inducing an increase in the average velocity in \eqr{rvzp}. As seen in \figr{cb} top panel, the RV increase at mid transit is larger than the RV decrease close to transit start and end. This reflects the geometry of the model and the limb darkened stellar surface. The maximum increase in RV can be approximated as:
\begin{equation}
\label{eq:acb}
A_{\rm CB} \approx  f V_{\rm CB} \frac{r^2}{1-r^2}\ ,
\end{equation}
where $f$ is determined by the integrals in \eqr{f} and $r$ is the planet to star radii ratio. For the example presented in \figr{cb}, $A_{\rm CB} = 2.3$ \ms. We caution that \eqr{acb} is a rough estimate, and $A_{\rm CB}$ decreases for inclined orbits with $i\ \textless\ 90$ deg. An order of magnitude comparison to the RM RV amplitude \citep[][Eq.~6]{gaudi07}, gives:
\begin{equation}
\label{eq:comp}
\frac{A_{\rm CB}}{K_{\rm RM}} \approx  \frac{V_{\rm CB}}{V_{rot}}\ .
\end{equation}
This ratio is 25 \% for the example presented in \figr{cb}, which uses $V_{rot}$ = 2,000 \ms, and it is larger for stars rotating more slowly. A more accurate estimate of the above ratio can be achieved by multiplying \vcb\ by $f$ in the right hand side of \eqr{comp}, resulting in a ratio of 18 \% for our adopted system, while the numerical integration gives a ratio of 12 \%.

The CB-induced RV curve does not depend on $\lambda$ or $V_{rot}$, meaning that on its own it does not hold information about the system's spin-orbit alignment. In fact, the CB RV signal should be present, in principle, even if the host star does not rotate at all. Since the measured RV curve includes both effects, the CB and RM, not accounting for the CB effect in the fitted model may result in a small bias of the spin-orbit angle. This is especially true for bright host stars that are also quiet, meaning show a low stellar activity level, as those stars allow a measured  RV precision comparable to and even below the amplitude of the CB-induced RV signal.

\begin{figure}
\begin{center}
\includegraphics[scale=0.70]{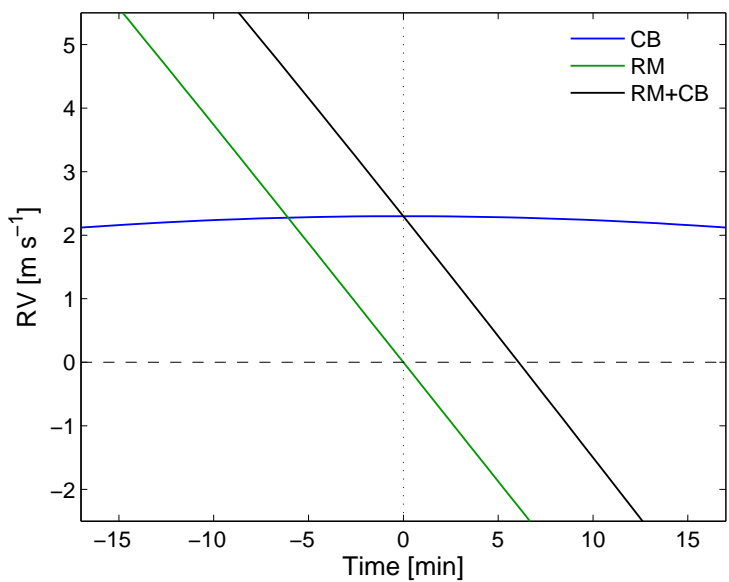}
\caption{\label{fig:cbzoom} 
Zoomed-in view of Fig.~1 middle panel around mid transit time. The different RV curves are represented using the same color code as in Fig.~1. The figure shows that for orbits close to edge-on or spin-orbit aligned, while the RM RV effect (solid green line) briefly cancels out and crosses the zero point (dashed black line, which is equivalent to the Keplerian signal) at mid transit time (vertical dotted black line), the RV curve including the RM and CB effects (solid black line) crosses the Keplerian signal at a later time. This time difference is quantified in \eqr{timediff} and equals 6 min in this example. The blue solid line is the RV curve induced by the CB effect.
}
\end{center}
\end{figure}

The symmetric shape of the RV curve induced by the CB effect distorts the shape of the RM signal. Specifically, for completely edge-on orbits (not necessarily aligned) or aligned orbits (not necessarily edge-on) the RM curve is an {\it antisymmetric} function about the mid transit time, so the combined RM + CB curve will have no definite symmetry, and the CB RV curve can not be completely absorbed into an RM model simply by changing $V_{rot}$ or $r$. Therefore, the CB effect can bias the interpretation of the RV curve measured during transit and the estimate of $\lambda$. For orbits close to edge-on the distortion is strongest during the central part of the transit where it increases the measured RV. \figr{cbzoom} presents a zoomed-in view of the middle panel of \figr{cb} around the mid transit time. It shows that the CB + RM RV curve (black solid line) crosses the RV zero point (dashed black line), meaning it equals the Keplerian velocity, later than when the RM curve does, which is at the mid transit time (dotted vertical black line), when the RM effect briefly cancels out. This time difference is approximated by dividing the CB signal at mid transit time ($A_{\rm CB}$) by the slope of the RM RV curve:
\begin{equation}
\label{eq:timediff}
\Delta t \approx f V_{\rm CB} \frac{r^2}{1-r^2} \ / \ \frac{2V_{rot}\frac{r^2}{1-r^2}}{\tau} = f \frac{V_{\rm CB}}{V_{rot}} \left( \frac{P}{\rm day} \right)^{1/3} \left( \frac{R_s}{R_{\sun}} \right) \left( \frac{M_s}{M_{\sun}} \right)^{-1/3}\ 54.4\ {\rm min}\ ,
\end{equation}
where $\tau$ is the duration of the full transit\footnote{The time from the second to third points of contact \citep{seager03}, when the planet's disk is completely enclosed by the stellar disk.}, $P$ is the orbital period, and $M_s$ is the host star mass. \eqr{timediff} shows that the above time difference can easily be a few minutes or more. It is 6 min for the system presented in \figr{cb}. Of course this time difference is dependent on the inclination angle, and for systems far from edge-on it will depend on $\lambda$ as well. However, for systems close to edge-on it will depend only weakly on $\lambda$, unless $\lambda$ is close to 90 or 270 deg. For example, for the system simulated here the above time difference is mimicked by the RM effect alone by increasing the impact parameter to about 0.25 and taking $\lambda \approx -15$ deg.

Therefore, an accurate estimate of the mid transit time {\it and} the expected Keplerian RV curve during transit are important for the correct interpretation of the spectroscopic transit RV curve and the identification of the CB effect. 
An accurate mid transit time can be obtained through simultaneous photometry --- the measurement of the photometric light curve of the same transit event observed spectroscopically \citep{hirano10, winn11} --- since transit light curves are known to provide mid transit time estimates with an accuracy of 1 min and better. The required uncertainty on the Keplerian RV curve during transit is smaller than the CB effect amplitude. Such a low uncertainty on this RV curve can be achieved by obtaining RVs out-of-transit, both right before and right after the transit, and also on adjacent nights in orbital phases further away from the transit, to allow for an accurate orbital characterization, including specifically the RV slope during transit and the system's center of mass velocity.

In addition, we note that for edge-on orbits with $\lambda$ of 90 or 270 deg the RM signal vanishes and the in-transit RV curve will be dominated by the CB effect. 

As we already mentioned the model we used here is probably a simplified version of the true behavior of the CB effect. Specifically, it was noted by several authors \citep[e.g.,][]{beckers78, howard80} that the dependency of the observed local blueshift velocity scales as $\mu^\alpha V_{\rm CB}$, where $\alpha > 1$. This leads to a faster decrease in the observed Doppler shift towards the stellar limb, faster than in the model we adopted here where $\alpha = 1$. The reason for this is the longer optical path in the stellar atmosphere closer to the limb. Using models with $\alpha > 1$ in our calculations results in increasing $A_{\rm CB}$.

Although beyond the scope of this paper, it is worth noting that the CB effect can influence the analysis of the RM effect in stellar binaries,  where the latter is measured through a careful analysis of the variation in the spectral lines shape \citep[][and references therein]{albrecht11}.

\section{Summary}
\label{sec:sum}

We have presented here the CB effect and its impact on the RV curve during a planetary transit. We adopted a simplified model where every surface element is radiating with the same blueshift velocity, directed along the normal to the surface. Using numerical integration across the observed stellar surface we calculated the RV curves for an imaginary transiting star-planet system, similar to some of the well known transiting systems with a close-in giant planet.

The amplitude of the anomalous signal induced by the CB effect during transit is at the 1 \ms\ level for giant transiting planets. Not accounting for the CB effect can lead to inaccurate interpretation of the RV curve, possibly impacting the estimate of the spin-orbit angle, which will be especially visible in the case of bright and quiet host stars with accurate RVs. 

For close to edge-on orbits the CB RV signal during transit is strongest at mid transit time, where it causes a time shift between the mid transit time and the time when the RV curve crosses the Keplerian curve. Therefore, an accurate estimate of the mid transit time and a well determined orbit are important for an accurate analysis of the spectroscopic transit.

So far the CB effect has been studied in detail only for the Sun. Future high-precision RVs will allow us to study it also for bright stars orbited by transiting planets, showing yet another way through which transiting planets can be used to probe their host stars. A measurement of the CB velocity for a large sample of stars will help to better understand processes such as convection and granulation, which in turn will help to better determine absolute RVs.

\acknowledgments

We are grateful to {\AA}ke Nordlund for useful discussions, and to Dainis Dravins and Eric Agol for important comments.
AS acknowledges support from NASA Grant Number NNX10AG02A.

 



\clearpage

\begin{deluxetable}{lcc}
\tablecaption{\label{tab:params}
Table of parameters of the transiting planetary system used in the example presented in Figs.~1 and 2.
 } 
\tablewidth{0pt}
\tablehead{\colhead{Parameter} & \colhead{Value} & \colhead{Units}  } 
\startdata
\multicolumn{3}{l}{Orbital parameters:} \\
Orbital Period, $P$ 			& 3.5		 	& day  \\
Inclination, $i$				& 90			& deg  \\
Eccentricity, $e$			& 0			& --- \\
Spin-orbit angle, $\lambda$ 	& 0			& deg  \\
\hline
\multicolumn{3}{l}{Host star parameters:} \\
Rotational velocity, $V_{rot}$	& 2,000		& \ms  \\
Convective blueshift, \vcb		& $-500\ $		& \ms  \\
Star radius, $R_s$			& 1.10		& \rsun  \\
Star mass, $M_s$			& 1.05		& \msun  \\
\hline
\multicolumn{3}{l}{Planet parameters:} \\
Planet radius, $R_p$		& 1.2			& $R_J$ \\
Planet mass, $M_p$			& 0.8			& $M_J$  \\

\enddata
\end{deluxetable}

\end{document}